# Application of Machine Learning Based Pattern Recognition in IoT Devices: Review

Zachary Menter[1], Wei Zhong Tee[1], Rushit Dave[1]

[1] University of Wisconsin-Eau Claire, Eau Claire WI 54701, USA

**Abstract.** The Internet of Things (IoT) is a rapidly advancing area of technology that has quickly become more widespread in recent years. With greater numbers of everyday objects being connected to the internet, many different innovations have been presented to make our everyday lives more straightforward. Pattern recognition is extremely prevalent in IoT devices because of the many applications and benefits that can come from it. A multitude of studies have been conducted with the intention of improving speed and accuracy, decreasing complexity, and reducing the overall required processing power of pattern recognition algorithms in IoT devices. After reviewing the applications of different machine learning algorithms, results vary from case to case but a general conclusion can be drawn that the optimal machine learning based pattern recognition algorithms to be used with IoT devices are K-Nearest Neighbor, Random Forest, and Support Vector Machine.

## 1    Introduction

The Internet of Things (IoT) is a term used to describe the network of devices, or things, with embedded software, sensors, transmitters and receivers, and other technology. This network collects, sends, and receives data from other connected devices via the Internet[1]. The emergence of IoT has allowed for a multitude of innovations in many different areas such as home automation, event prediction, and activity recognition, to name a few. Nowadays, many complex calculations and machine learning algorithms that used to require large amounts of processing power can all be run on IoT devices, leading to many exciting and inventive applications[2]. The multitude of embedded sensors in many IoT devices allows for many innovations through pattern recognition in everyday life. These innovations in pattern recognition allow for our mobile and IoT devices to serve new and better functions every day. Because of this, the use of pattern recognition plays a massive role in a large majority of the present work being done in IoT devices. Pattern Recognition is the notion of assigning objects to classes. Such patterns that can be recognized include textures, images, speech, biological/physical features, habits and many other types of patterns. Features of the object are organized in a selected space where an algorithm of technique is used to assign it a class label[3].

Utilizing Machine Learning algorithms has become popular among IoT devices because it improves IoT based services such as traffic engineering[4], security[5-8] and security assessment[9], speaker[10] and image[11] recognition, and quality of service optimization[12]. Pattern recognition in IoT devices has also seen much more development and consumer use in smart home[13], automation[14] and cloud based monitoring and automation[15] systems. Algorithms must be chosen based on their power efficiency and computational cost and their ability to correctly classify features as most IoT based systems would have constraints with power, memory or storage[16,17]. Efficient and intelligent IoT applications have emerged as a result of employing machine learning algorithms into IoT based devices and services.

The combination of pattern recognition and IoT devices has been deployed in many industries and are often developed on a generic basis[18], allowing for it to be used in many different situations. Various industries have utilized pattern recognition and IoT devices for different causes such as strengthening security and implementing biometric solutions, recognizing vehicle/traffic patterns, predicting complex events, recognizing and classifying human activity and automating tasks.

## 2    Background

There is great significance in selecting the best machine learning algorithm for pattern recognition in IoT devices. Given the restraints IoT devices have with memory, storage, or power consumption, an efficient algorithm will allow for fast processing time, low space usage and low power consumption. Existing work that is based on Human Activity Recognition (HAR) and Biometric Security has been developed and surveyed for this article.

Pattern recognition in IoT devices for human activity recognition has seen such an increase in demand and development because the devices and sensors that are capable of recording human motion or the vital signs data have to be light, compact and wearable[19]. Most commonly, predefined activity models are first used to train classifiers to identify activities performed by humans based on data collected by various wearable sensors[20]. A study carried out by Shwet Ketu and Pramod Kumar Mishra[21] in 2020 published a performance analysis of machine learning algorithms for IoT based Human Activity Recognition. Different algorithms consist of different capabilities and performance, and in this case, factors such as the run time, space required, and energy consumed were measured with the aim of selecting the most optimal algorithm for wearable sensors for human activity recognition using a predefined activity recognition dataset. This process included selecting various algorithms and running test cases through them in a virtual simulation that evaluates its accuracy, precision, recall and F-1 score. The testing concluded that seven of the fourteen algorithms that were selected performed better with higher accuracy; hence using one those seven algorithms for human activity recognition under similar circumstances would be more optimal than the others. The seven algorithms include the Gradient Boosting Classifier, Random Forest, Bagging Classifier, Classification and Regression Trees, Support Vector Machine, K Nearest Neighbor and the Extra Trees Classifier. Similarly, another common use of machine learning algorithms in IoT devices for pattern recognition is in security systems. One type of security that IoT devices handle is biometrics, which includes recognizing faces or fingerprints[22]. Security is also very important in smart home technology. IoT devices are the main components of every smart home setup. Smart devices generally have wireless access to a user's accounts and home devices. This produces a large need for effective, compact security solutions. These IoT devices are at risk for a number of threats such as information leaks, data mining, denial of service attacks, and various other cyber-attacks[23]. One of the ways these attacks can be detected and prevented is through statistical analysis and machine learning, which can help inspect and detect anomalies in the data being sent over a network[24]. Cong Shi, Jian Liu, Hongbo Liu, and Yingying Chen[25] developed a spoofer detection model using the Support Vector Machine algorithm. This model was able to correctly identify between legitimate users and spoofed users. In addition, much work has been conducted on applying various machine learning methods to intrusion detection systems. With many new models and developments being made frequently, it is important to review and analyze the work that has been done in order to seek out possible and necessary improvements. Though much has been done in the way of machine learning in human activity recognition and security in IoT devices, each model and method used may have certain advantages and disadvantages. With this study, we seek to explore these advantages and disadvantages to improve the work that is being done in this field.

## 3 Human Activity Recognition (HAR) in IoT Devices

### 3.1 Opportunistic Sensing for Inferring in-the-wild Human Contexts Based on Activity Pattern Recognition Using Smart Computing

A new approach for activity-aware human context recognition (AAHCR) using both a smartphone and smartwatch together to infer the user's context information based on pattern recognition was introduced in this study. This is known as activity aware human context recognition (AAHCR). Daily Living Activities (DLAs) used in the proposed scheme include lying down, running, sitting, standing and walking.

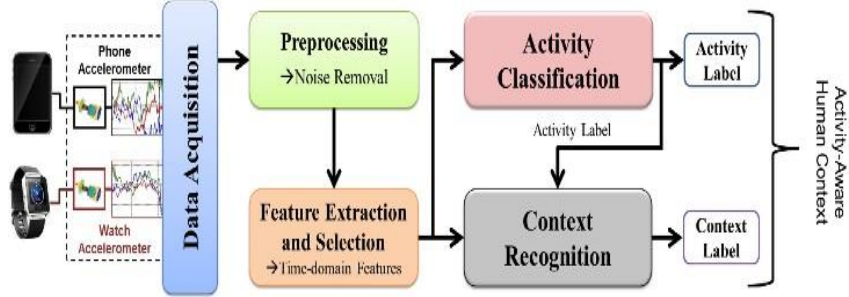

**Fig. 1.** Proposed methodology for the model that is AAHCR based[26].

The methodology of the study is shown in figure 1. The data that was used is a publicly available dataset called the Extrasensory dataset. It is from a different study that collected data from 60 participants regarding in-the-wild activities. The Extrasensory dataset contained binary context labels that corresponded to the selected DLAs. The two context labels that each activity consisted of were the behavioral contexts and the phone positions. Such positions include the phone on a table, in a hand, pocket or bag. The context information that concerns the user's secondary activity is incorporated into the first layer of determining the DLA. The second layer represents the phone's context when a specific physical activity is performed, which the first layer utilizes for position dependent human context recognition. A low-cost time domain soothing filter was chosen for signal denoising. Signal attributes are determined by the feature extraction process. The extracted features are then used for activity classification and the context recognition. Two types of context recognition experiments were conducted: position independent and position dependent. In position independent, no phone position information is incorporated into the classifiers training. In position dependent, the classifiers are trained for different phone positions. The selected pattern recognition put were compared in the model includes Decision Tree (DT), Random Forest (RF), K-Nearest Neighbors (K-NN), Bayes Net (BN) and Multilayer Perceptron (MLP-ANN). In both the position dependent and position independent tests, the RF classifier showed the best results in the majority of metrics (precision, recall, F-1 score, balance accuracy, kappa and root mean squared error) when it was paired with the data collected from both the smartphone and watch accelerometer sensors. However, the position dependent tests offered better results from the metrics compared to that of the position independent tests. As the RF classifier was determined to be the ideal algorithm for this model, a confusion matrix was then generated to represent the predicted DLAs from a fusion of data from both sensors. Static activities such as lying down, sitting and standing have a high percentage of correct predictions whereas dynamic activities such as walking or running had a slightly lower percentage of correct predictions. The study concluded that different human behavioral contexts can be used to recognize and predict daily activities. The data provided by Extrasensory showed that the best algorithm for pattern recognition in human activities using behavioral contexts was the Random Forest.

## 3.2 Wearable-Based Human Activity Recognition Using an IoT Approach

A study conducted in 2017 proposed using a remote monitoring component with remote visualization and programmable alarms for Human Activity Recognition (HAR) and validates the approach used. It is made of two main components: a traditional Human Activity Recognition (HAR) system that can be used on any mobile or non-mobile device and an application for recognizing and surveilling in a health care related subject. The flow of the model is shown in figure 2. The learning phase establishes the relations between the data and activities. It first collects the data from all the sensors the system is using. Sensors will be dependent on the kind of device that is being built for recognition. Time, the type and duration are the factors that need to be recorded in an activity log and the activities should be carried out in a random order with random durations. The feature extraction step is then based on structural features and statistical features. Structural features often fit in a defined mathematical function and there is no correlation between the signals in the data. Statistical features extract its features based on the statistical information such as the mean, standard deviation or correlation. Lastly, the learning phase develops a recognition model that uses the data set, activity log and features to recognize activities. Following the learning phase, the first step in the recognition phase is the data collection. However, it does not use any prior knowledge on the activities performed so it does not keep an activity log. Similarly, it extracts features in the same fashion as the learning phase using structural and statistical features. The Zephyr Bio harness 3 was chosen as the wearable tracker because of its capability of measuring the required variables and connecting to a smartphone via Bluetooth. A smartphone application is needed to receive the data and handle the communication and storage of the raw data of the wearable. The cloud component receives all the raw data of the wearable and the activities recognized by the HAR classifier. A platform called Ubidots was selected. It could visualize the history of the activities recognized, heart rate, respiration rate, posture and the acceleration values. The data that is used is the heart rate, respiration rate, posture, three-axis acceleration, peak acceleration and electrocardiogram magnitude. The structure detection algorithm searches for the best fitting mathematical function for groups of data. The training data set for the classifier included 14 samples of weather features, with no statistical correlation between instances. With this, each algorithm generated its own set of rules. The resulting rule sets were a C4.5 algorithm and a Naive Bayes algorithm. The selected classifier was the C4.5 algorithm as it was efficient and used far less space than that of the Naive Bayes algorithm. After proving that it worked with the weather samples, the model was implemented to work with training data to recognize . After the feature extraction phase, the algorithm proceeds generating the rule tree to that specific training data set. In total, there were 13 rules for having a single recognition model for the different subjects that were involved during this phase. 69 out of 72 tasks were successfully recognized for one of the random test subjects. The system correctly classified test subjects who were lying down or jogging. The few errors made by the system were when it had to classify whether a test subject was sitting or walking. It was then concluded that a human activity recognition system using a smartphone, Bio harness and cloud system could be successfully developed and implemented.

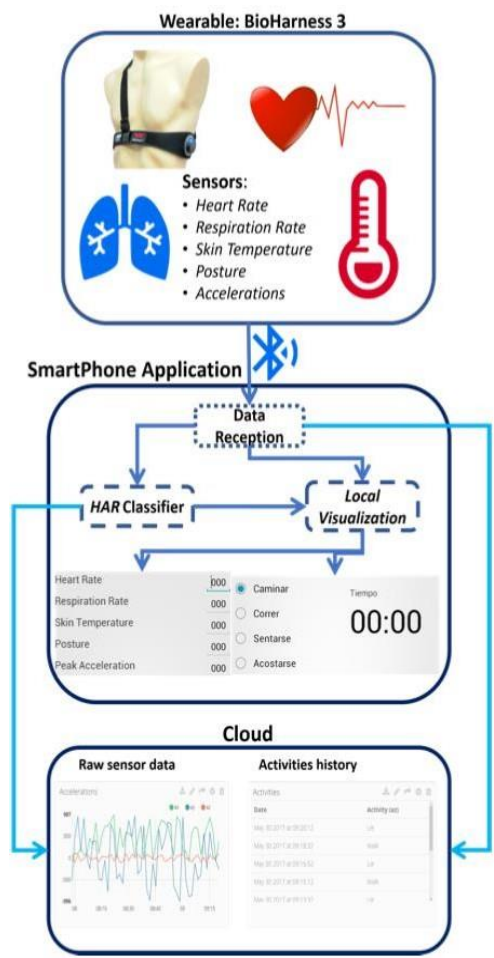

**Fig. 2.** Proposed flow of the model generation and activity recognition phases[27].

### 3.3    Generalized Activity Recognition Using Accelerometer in Wearable Devices for IoT Applications

One of the most common implementations of HAR systems includes generalized activity recognition model for wearable devices. A diagram of such a model is shown in figure 3. It covers how the automatic detection of different activities work using just one axis in an accelerometer and the simple features and pattern recognition algorithms that were used that were effective, computationally inexpensive and suitable for wearable devices with constrained resources.

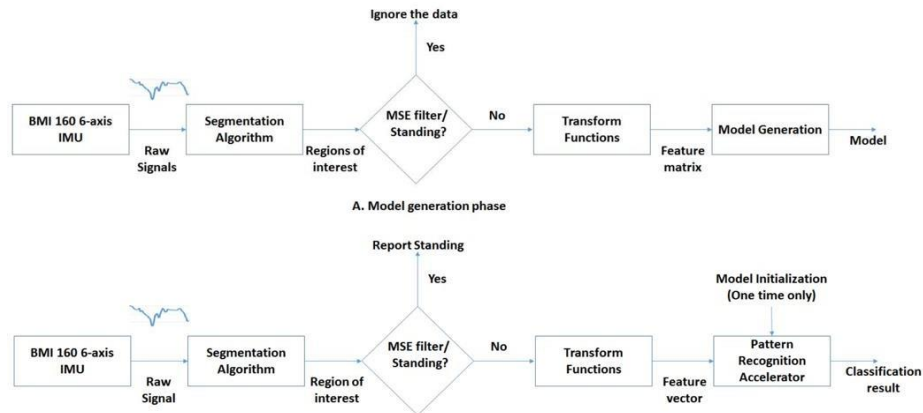

**Fig. 3.** The proposed flow of the model generation and activity recognition phases[28].

Data was collected in a custom-built device that contained Bosch's BMI 160. The devices were positioned on the sacrum of each test subject and it sampled at a rate of 100Hz. Data from each subject was collected while they were performing tasks that included walking, running, crawling, ladder climbing and pronating for 30 seconds each. The classification model was trained from data collected from 52 male subjects in firefighting gear and tested on different times from the set of users. Test data was also obtained from three separate sets from eight subjects, of which there was a mixture of male and female in normal street clothing test subjects. The recognition system only uses data from the Y axis of the accelerometer for the segmentation, feature extraction and recognition. A mean square error (MSE) filter was used to the segmented data to determine if the test subject was standing. A supervised hierarchical clustering algorithm was used to cluster the feature vectors based on similarity characteristics and labels. A centroid was computed for each cluster and each centroid was assigned a label based on majority. The pattern recognition capability was then leveraged by optimizing the number of centroids to use a finite surface around each centroid. To select the final model, multiple iterations with new feature vectors from all but one subject from the training set and left out feature vectors form the test set. In each iteration, a new model is developed, and its performance tested on the test set using metrics such as recall, precision and accuracy. In the recognition phase, each test vector is assigned a class based on its similarities to patterns. In the event the test vector does not match any or falls out of the surface of the pattern, it is assigned an unknown classification (UNK). The decision tree model was only trained using training data that consisted of walking and climbing data in order for the model to be able to distinguish the difference of the two classes. To prevent the issue of overfitting, the tree was pruned by varying the tree depth. A decision tree with a depth of three resulted in the best validation performance. The first test consisted of 66 patterns selected based on the datasets that contained 52 subjects. It got an average f1-score of 0.91 and it struggled to identify crawling and pronating the most. The second test was carried out on eight subjects with a mix of males and females and street clothes. The average f1-score was 0.88. The final test also was carried out on eight subjects and the test ended with an average f1-score of 0.91. The model showed accurate results with detecting walking and running but did not see the same success with recognizing crawling.

### 3.4 A Platform and Methodology Enabling Real-Time Motion Pattern Recognition on Low-Power Smart Devices

Another study presented a low powered smart device known as the Neblina system on modules with hardware variants and expansion modules that targets IoT applications. The accuracy, performance, memory and power consumption of the Neblina are actively monitored when implementing and testing the proposed Motion Pattern Recognition (MPR) on a fitness activity data set.

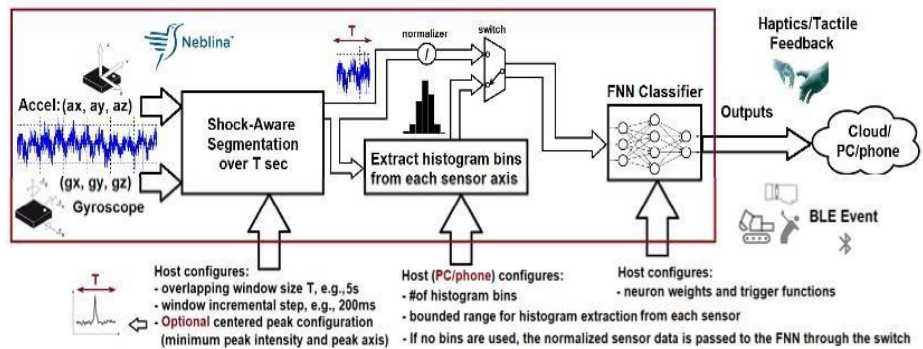

**Fig. 4.** A figure caption is always placed below the illustration. Short captions are centered, while long ones are justified. The macro button chooses the correct format automatically[29].

The Neblina consists of low power sensors, on board memory and Bluetooth Low Energy. The base hardware has an efficient 9 axis orientation tracking algorithm, using its magnetic sensors and a framework, shown in figure 4, using inertial sensors. The framework contains the shock aware segmentation, feature extraction and classification. A 3 axis accelerometer and gyroscope is what makes the shock awareness segmentation possible. The sensors have all passed the factory's calibration and filtering techniques. It reads the acceleration that the sensors pick up and compares it to peak constraints to detect specific activities. A segmentation process is based on an overlapping sliding window with the accelerometer and gyroscope sensors. The sensors would have needed to pass factory calibration techniques. The acceleration magnitude is used to determine the minimum peak intensity in the motion segment characterization. The feature extraction uses a time domain histogram instead of more conventional methods. This is because when created with high enough bins, a time domain histogram gives accurate distributions. It is also low in terms of latency and computational costs, making it ideal for real-time execution. The classification is performed using a single hidden layer Feedforward Neural Network (FNN), of which the weights and trigger functions are set by the host. It is favorable because the FNN can derive arbitrary nonlinearities from its inputs and it is also efficient with its latency and memory usage. The FNN is also proven to have better latency and RAM usage compared to the segmentation and histogram extraction and the more conventional KNN classifier. Only 10% of random data points collected were chosen and used to train the classifier. With 60 hidden neurons and 7 epochs, an accuracy of 99.6% was achieved during training. F1 scores for each activity were between 83% and 90% as well for the activities mentioned previously. Furthermore, the Neblina managed to last about 41 hours even with the small 100mAh battery, which proved that this was power efficient.

### 3.5 IoT Based Mobile Healthcare System for Human Activity Recognition

The final article that was reviewed presented a health care system known as mHealth that is based on IoT and mobile devices. This is part of the m-healthcare system which utilizes mobile devices and wearable body sensors. The MHEALTH dataset was used as it contained ten volunteer's vital sign recordings and body motions for several physical activities. Sensors were placed on each subject's left ankle, right wrist, and chest. The twelve activities that were recorded were standing still, sitting/relaxing, lying down, walking, climbing stairs, bending their waist forward, frontal elevation of arms, knees bending, cycling, jogging, running and jumping front and back. Different data mining techniques were used in the Human Activity Recognition (HAR). The sensor acquired data at rates of 50Hz. The activities were recorded without any constraints and outside of a laboratory with no controlled variables. The model's predictive ability was then put to the test with 10 subjects using eight different algorithms.

**Table 1.** The results of average classification accuracy (CA), F-measure (F-M) and area under the ROC curve (AUC) of the different Algorithms[30].

|  | k-NN | ANN | SVM | C4.5 | CART | Random Forest | Rotation Forest |
|---|---|---|---|---|---|---|---|
| Average CA | 66.64 | 99.55 | 99.89 | 99.32 | 99.13 | 99.89 | 99.79 |
| Average F-M | 0.997 | 0.996 | 1 | 0.99 | 0.991 | 0.9989 | 0.9979 |
| Average AUC | 1 | 1 | 1 | 0.998 | 0.998 | 1 | 1 |

Results of the experiment are shown in table 1. The average accuracy for the Random Forest and SVM were equal at 99.89%. However, since the Random Forest is faster than the SVM, it was chosen for HAR. The table below shows the average classification accuracy, F-measure and ROC for each data mining technique that was used 10 times for each subject while testing.

## 4 Security and Pattern Recognition in IoT Devices

### 4.1 Continuous Authentication of Smartphone Users Based on Activity Pattern Recognition Using Passive Mobile Sensing

A study conducted in 2018 used various machine learning classifiers to authenticate smartphone users unobtrusively and continuously by utilizing passive mobile sensing. Common unlocking methods fail to authenticate the user continuously, meaning that if a phone is unlocked, anyone can use it. This proposed method uses machine learning classifiers to detect and recognize physical activity patterns in smartphone users to provide continuous authentication.

In this study, the phone's accelerometer, gyroscope, and magnetometer were used. Six different activities were used for user identification: walking, running, standing, sitting, walking upstairs, and walking downstairs. The system was trained to learn behavioral patterns for different users on all six of the activities. The system was also trained on five different smartphone positions on the body. These positions were the upper arm, wrist, waist, right thigh, and left thigh. The proposed model is shown in figure 5.

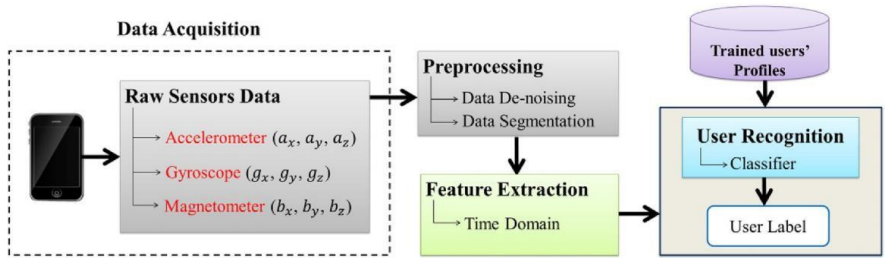

**Fig. 5.** The flow of the proposed model for user recognition[32].

The dataset used in this study is a publicly available dataset gathered by a previous study[32] performed on physical activity recognition. Initially, 16 different features that have proven effective in other studies were extracted from the dataset. In this study, they compared the performance of three different classifiers: Support Vector Machine (SVM), Decision Tree (DT), and K-Nearest Neighbors (KNN). Each of these classifiers were trained separately for different activity patterns of all ten participants in the study.

**Table 2.** Average performance of the classifiers with smartphone in the waist position[32].

| Classifier | Accuracy | Precision | Recall | F-measure | Error Rate |
|---|---|---|---|---|---|
| DT | .971 | .972 | .971 | .971 | .029 |
| KNN | .923 | .924 | .923 | .924 | .078 |
| SVM | .986 | .986 | .986 | .986 | .014 |

The performance of the chosen classifiers is shown in table 2. The SVM classifier on average outperformed both the DT and KNN classifiers with higher accuracy and precision and a lower error rate. One limitation of this system is that it can only identify users based on the activities that the system has been trained for. In the future, more sensors and activities can be added to the system to increase the accuracy and number of situations that the system will work in.

### 4.2 Please Hold On: Unobtrusive User Authentication Using Smartphone's Built-in Sensors

Common smartphone authentication methods, though simple to perform, are relatively time consuming and obstructive when performed multiple times a day. A study from 2015 sought to solve that problem by introducing an unobtrusive user authentication method based on the micromovements of the user's hands. This method was chosen so that the user can just swipe to unlock their phone and the phone will recognize the user based on their hand movements after unlocking.

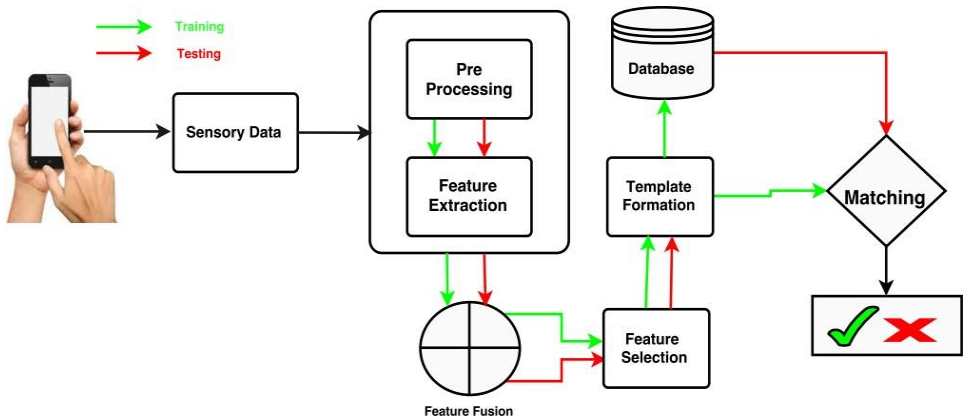

**Fig. 6.** The flow of the proposed model for user authentication[33].

The methodology, shown in figure 6, includes profiling the smartphone user's hand movements for a short time after unlocking the phone. To collect the required data, the researchers developed an Android app called Data Collector. They ran the experiment on data collected everywhere between two and ten seconds after unlocking the phone. Seven different statistical values were gathered for different time intervals from all four dimensions of each sensor's data. Therefore, 28 features were extracted from each sensor. Four different machine learning algorithms were chosen for classification: Bayes Net (BN), K-Nearest Neighbor (KNN), Multilayer Perceptron (MLP), and Random Forest (RF).

**Table 3.** Results of different classifiers for different lengths of data collection[33].

| Classifier | | 2s | 4s | 6s | 8s | 10s |
|---|---|---|---|---|---|---|
| BN | TAR | .89 | .89 | .89 | .88 | .89 |
| | EER | .11 | .11 | .11 | .12 | .12 |
| MLP | TAR | .93 | .93 | .94 | .94 | .94 |
| | EER | .07 | .07 | .06 | .06 | .06 |
| KNN | TAR | .88 | .88 | .89 | .89 | .90 |
| | EER | .12 | .12 | .11 | .11 | .10 |
| RF | TAR | .95 | .95 | .95 | .95 | .95 |
| | EER | .05 | .05 | .05 | .05 | .05 |

Random Forest and Multilayer Perceptron both yielded the best results with the default parameters, shown in table 3. MLP performed best on 10 seconds and 6 seconds of data collection whereas RF was consistent on all durations. One limitation of this model is that it does not consider the impact of different situations during authentication such as running, walking, standing, sitting, etc. These situations could influence the behavioral patterns being classified as the behavioral patterns of a person standing still may be different than the behavioral patterns of the same person while walking.

### 4.3 Selection of Effective Machine Learning Algorithm and Bot-IoT Attacks Traffic Identification for Internet of Things in Smart City

Another study from 2019 proposed a model and hybrid algorithm for selecting machine learning algorithms for cyber-attack traffic detection. Many organizations have the need for IoT threat detection, but it is not always clear which machine learning algorithm is best suited for the job. This study proposed a framework to solve this problem.

The dataset used in this study was the Bot-IoT dataset. 44 of the most effective features were selected from the dataset. The purpose of the system is to select the best machine learning algorithm for a problem from a set of machine learning algorithms. The set of algorithms that they chose to use were Naïve Bayes, Bayes Net, Decision Tree C4.5, Random Forest, and Random Tree. To find the most effective machine learning algorithm, they used a mathematical tool called the bijective soft set. This technique has been effective in multiple other studies related to decision making, so they opted to apply it to this as well. The bijective soft set algorithm, shown in figure 7, calculates soft sets for each machine learning attribute then for each of those soft sets, calculates the correlation AND OR product. After correlation, the algorithm calculates the union operation and the intersection operation to reduce the correlation table to 1x1. The proposed algorithm, shown in figure 8, is a hybrid machine learning algorithm selection algorithm for anomaly and intrusion detection in IoT networks. The proposed algorithm calculates the performance result values of each of the selected machine learning algorithms. The algorithm then ranks each machine learning algorithm based on threshold values. After this, the algorithm calculates the attributes and runs them through the bijective soft set algorithm.

| Algorithm 1: ML algorithm selection based on Bijective soft set: |
|---|
| Input: ML (ML$_1$, ML$_2$, ML$_3$, .... ML$_n$)   // set of ML algorithms, |
| Output: ML algorithm []   // selected effective algorithm |
| 1.    begin |
| 2.    for i = 1 to N   // ML attributes (metrics) |
| 3.        calculate soft set from each [i] for each attributes; |
| 4.    end for |
| 5.    for i = to N; |
| 6.        calculate Correlation both AND OR product n×n; |
| 7.        calculate Union operation either in row or column |
|            to reduce the correlation table to 1×n or n×1; |
| 8.        calculate Intersection operation either in row or |
|            column for possible reduction the correlation table |
|            to 1×1; |
| 9.    end for |
| 10.    Finally, obtain the effective ML algorithm, if the table is |
|        Reduced to 1×1; |
| Return Algo; |

| Algorithm 2: ML algorithm selection based on Bijective soft set: Combined with selected algorithms: |
|---|
| Input: MLA (M$_1$, M$_2$, M$_3$,.... M$_n$)        // ML algorithm set, |
| Output: ML algorithm []                    // selected ML algorithm |
| 1.    begin |
| 2.    for i = 1 to M |
| 3.        calculate value [i] for each algorithm; |
| 4.    end for |
| 5.    for i = 1 to N; |
| 6.        calculate ranking (Fi); |
| 7.        if (ranking(F) > δ); |
| 8.            Insert Fi into descending order; |
| 9.        end if |
| 10.    end for |
| 11.    for i = 1 to N   // ML attributes (metrics) |
| 12.        calculate soft set from each [i] for each attributes; |
| 13.    end for |
| 14.    for i = to N; |
| 15.        calculate Correlation both AND OR product n×n; |
| 16.        calculate Union operation either in row or column |
|            to reduce the correlation table to 1×n or n×1; |
| 17.        calculate Intersection operation either in row or |
|            column for possible reduction the correlation table |
|            to 1×1; |
| 18.    end for |
| 19.    Finally, obtain the results of bijective soft set, if the table is |
|        Reduced to 1×1; |
| Return Algo; |

**Fig. 7.** The bijective soft set algorithm[34].    **Fig. 8.** The proposed algorithm[34].

**Table 4.** Results for each of the applied machine learning algorithms[34].

| Algorithm | Accuracy | Precision | Recall | TPRate | TTBM |
|---|---|---|---|---|---|
| Naïve Bayes | 99.79 | 0.99 | 0.98 | 0.99 | 4.03 |
| Bayes Net | 99.77 | 1.00 | 0.99 | 0.99 | 29.26 |
| Decision Tree C4.5 | 99.99 | 1.00 | 1.00 | 1.00 | 17.1 |
| Random Forest | 99.99 | 1.00 | 1.00 | 1.00 | 198.83 |
| Random Tree | 99.99 | 1.00 | 1.00 | 1.00 | 4.32 |

The results of the applied machine learning algorithms are shown in table 4. It can be seen that all of the machine learning algorithms performed very well. However, Naïve Bayes and Random Tree stood out as the most effective algorithms. This is because, though all were very accurate, Naïve Bayes and Random Tree both had a very low time taken to build the model (TTBM). The researchers mentioned that though they measured the performance using the five metrics in table 4, the most important of these metrics are the accuracy and TTBM. This study is a good start with this method, but it is stated that they would like to try it with a larger set of machine learning algorithms, and in different scenarios other than anomaly and intrusion detection.

### 4.4 ProFiOt: Abnormal Behavior Profiling (ABP) of IoT Devices Based on a Machine Learning Approach

A study conducted in 2017 sought to build the abnormal behavior profiling of IoT devices using machine learning. Abnormal behavior profiling is important especially in IoT devices because of the wide range of device types and functions. Two different scenarios were tested, one where one piece of data from a sensor was faulty, and one where all of the pieces of data were faulty. This study helped demonstrate how a small modification in sensed data can affect a machine learning algorithms detection accuracy.

The proposed system was of a smart building with heating, ventilation, air conditioning, and a fire alarm. The sensors in the building would measure and report the temperature, humidity, light level, and voltage to a server. The proposed threat is of a hacker compromising one of these sensors through a malicious attack. The data used in this study was from the Intel Berkeley Lab. One sensor was chosen, and a profile was built for detecting abnormal behavior using the four attributes of temperature, humidity, light, and voltage.

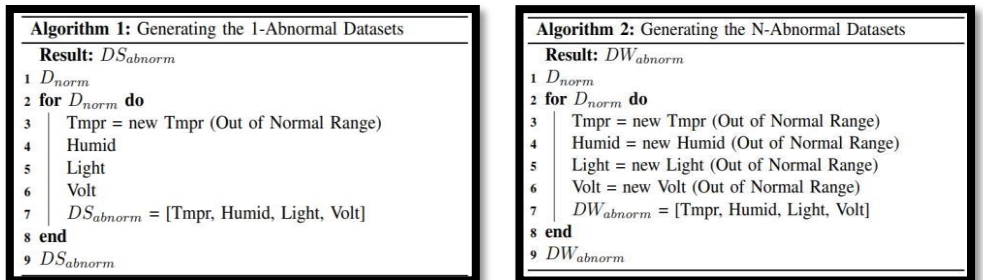

**Fig. 9.** The algorithms for dataset generation[35].

Two different abnormal datasets were generated, shown in figure 9. The first had only one attribute of the sensor modified. The second abnormal dataset had all attributes of the sensor modified. The training set was then fed into both a k-Means algorithm, and a Support Vector Machine algorithm.

**Table 5.** Detection rate using k-Means algorithm[35].

| Num of Clusters | 2 | 4 | 6 | 8 | 10 |
|---|---|---|---|---|---|
| 1-Abnormal | 78.3% | 88.8% | 93.1% | 95.0% | 97.0% |
| 4-Abnormal | 38.4% | 63.9% | 77.1% | 88.1% | 92.7% |

The k-Means algorithm, shown in table 5, worked well at detecting anomalies in the 1-abnormal dataset but performed worse on the 4-abnormal dataset. Conversely, the Support Vector Machine algorithm performed worse on the 1-abnormal dataset than it did on the 4-abnormal. They chose to use the k-Means algorithm for their abnormal behavior profiling system based on these results because it had the best overall performance on both datasets. This study shows that which machine learning algorithm you select for threat detection is important and may vary based on the situation and system that one is looking to implement.

## 5    Discussion and Analysis

**Table 6.** Methodology Analysis

| Title | Methodology | Results | Advantages and Disadvantages |
|---|---|---|---|
| Opportunistic Sensing for Inferring In-The-Wild Human Contexts Based On Activity Pattern Recognition Using Smart Computing | A model was built that used behavioral contexts to recognize human activity. A smartphone and watch were used for data collection and RF classifier was used for the algorithm in the model. | The RF classifier showed the best results across the metrics used in validating the most optimal algorithm. | Data collection was performed by easily accessible devices and results were accurate. The predictive model was limited to predefined contexts and could only predict a limited number of human activities. |
| Wearable-Based Human Activity Recognition Using an IoT Approach | This model uses remote monitoring component with remote visualization and programmable alarms for Human Activity Recognition (HAR). | The C4.5 algorithm was the optimal algorithm as it was efficient and used far less space than that of the Naive Bayes algorithm. | While many tasks were successfully recognized, the model struggled with classifying the difference between sitting and walking. |
| Generalized Activity Recognition Using Accelerometer in Wearable Devices for IoT Applications | Only one axis from accelerometer data is used to generate a computationally inexpensive recognition model. A supervised hierarchical clustering algorithm was used to cluster the feature vectors. | Multiple tests were carried out with different variables and the model had f-1 scores of 0.91, 0.88 and 0.91 and respectively. | The model showed accurate recognition capability with different environmental variables that were controlled during the tests. However, it struggled to identify crawling and pronating movements. |
| A Platform and Methodology Enabling Real-Time Motion Pattern Recognition on Low-Power Smart Devices | A low powered smart device known as the Neblina was used in a motion pattern recognition system based on fitness activity. | The FNN classifier proved to have better latency and RAM usage compared to the segmentation and histogram extraction and the more conventional KNN classifier. | Pattern recognition using the Neblina was accurate and incredibly power efficient, as it managed to operate for 41 hours with just a small 100mAh battery. |
| IoT Based Mobile Healthcare System for Human Activity Recognition | MHEALTH data was used to build a recognition model and multiple sensors were placed on subject's body for the human activity recognition. | The average accuracy for the Random Forest and SVM were equal at 99.89%. However, since the Random Forest is faster than the SVM, it was chosen for HAR. | The model that was developed was fast and accurate using a widely utilized data set. |
| Continuous Authentication of Smartphone Users Based on Activity Pattern Recognition Using Passive Mobile Sensing | Utilizing a publicly available dataset, user's behavioral patterns were classified and used for recognition and continuous authentication for smartphones. | Out of all tested classifiers, SVM performed best on average in all measures. | The proposed system allows for continuous user authentication rather than the standard one-time passcode unlock. The system only works with activities that it has been trained on, so any unrecognized activity will reject the user. |
| Please Hold On: Unobtrusive User Authentication Using Smartphone's Built-in Sensors | A new authentication method for unobtrusively authenticating smartphone users based on micromovements of the user's hands was introduced. Using collected data, four different classifiers were chosen for the model. | The Random Forest and Multilayer Perceptron performed the best. | This method allows a user to be authenticated unobtrusively, rather than having to input some type of password. The system does not consider the impact of different situations while authenticating. |
| Selection of Effective Machine Learning Algorithm and Bot-IoT Attacks Traffic Identification for Internet of Things in Smart City | Using the bijective soft set algorithm and the Bot-IoT dataset, a system was built to help select the best classifier out of a set of classifiers for the selected dataset. | All classifiers performed well, but Naïve Bayes and Random Tree had the lowest TTBM, thus making them the most effective. | The bijective soft set has been used for similar problems and was effective at finding a solution. However, the testing set of classifiers was relatively small. |
| ProFiOt: Abnormal Behavior Profiling (ABP) of IoT Devices Based on a Machine Learning Approach | Using data from the Intel Berkeley Lab, a system was built to detect abnormal behavior in IoT devices using the K-Means and Support Vector Machine algorithms. | Both algorithms performed better than the other in different situations. | The system was effective at detecting abnormal behaviors. From the algorithms that were used, there isn't a single best algorithm that can be used for every case. |

An analysis of the articles reviewed is shown in table 6. All the articles that were reviewed under the category of human activity recognition showed that different algorithms suited different methodologies accordingly. The type of data used, features extracted and power efficiency expectations play crucial roles in choosing the most optimal algorithm. Two articles that were surveyed involved user authentication using either behavior patterns or hand movements. Another article reviewed introduced a system for selecting the most effective machine learning algorithms. The final article for security in IoT devices involved recognizing and profiling IoT devices behaving abnormally.

### 5.1 Limitations

Some limitations of the models used in human activity recognition models that were reviewed was that the models were only limited to recognize activities based on the training data set and the feature extraction process. Should there be any new type of activity that the model would want to recognize, the model would have to be retrained with the new data containing the desired activities to re-extract the features again in order to build a working classifier. Furthermore, new parameters would have to be defined for the model generation (feature extraction) which contains the new activity that is being introduced in the model. In the security in IoT devices section, limitations on user authentication include the system not being trained on all activities or situations that may be encountered. This can lead to being locked out of the device when an activity or situation is not recognized. A limitation for the method for the selection of the best machine learning algorithm using the bijective soft set method is that the study was done on a relatively small set of algorithms. Similarly, a limitation for the study done on abnormal behavior profiling of IoT devices is that only two different algorithms were tested, meaning that there may be a superior algorithm that was not tested.

## 6 Conclusion

This article covered the use of pattern recognition in IoT devices for human activity recognition and security models. The literature conducted has proven that there are more optimum algorithms depending on the use case, data collection and design structure of the model. However, based on the literature reviewed, there are certain algorithms that stand out as the most effective when dealing with IoT devices due to their simplicity, time taken to build models, and accuracy. Different goals and expectations for the model influence the final selection of the optimal algorithm. However, based on the 10 articles that were reviewed, the most popular algorithms that were selected in models that consisted of pattern recognition in human activity recognition and security in IoT devices were the K-Nearest Neighbor, Random Forest, and Support Vector Machine.

# References

1. Atzori, L., Iera, A., Morabito, G.: The Internet of Things: A survey, vol. 54, pp. 2787-2805 (2010).
2. Mahdavinejad, M., Rezvan, M., Barekatain, M., Adibi, Peyman., Barnaghi, P., Sheth, A.: Machine learning for internet of things data analysis: a survey, vol. 4, pp. 161-175 (2018).
3. Kucheva, L., Whitaker, C.: Pattern Recognition and Classification (2015).
4. Akbar, A., Carrez, F., Moessner, K., Zoha, A.: Predicting complex events for pro-active IoT applications, pp. 327-332 (2015).
5. Dean, A., Agyeman, M, O.: A Study of the Advances in IoT Security (2018).
6. J. Shelton et al., "Palm Print Authentication on a Cloud Platform," 2018 International Conference on Advances in Big Data, Computing and Data mining.
7. Mason, J., Dave, R., Chatterjee, P., Graham-Allen, I., Esterline, A., & Roy, K. (2020, December). An Investigation of Biometric Authentication in the Healthcare Environment. Array, 8, 100042. doi:10.1016/j.array.2020.100042
8. Gunn, Dylan J. et al. "Touch-Based Active Cloud Authentication Using Traditional Machine Learning and LSTM on a Distributed Tensorflow Framework." *Int. J. Comput. Intell. Appl.* 18 (2019): 1950022:1-1950022:16.
9. Sundaram, K., Swarup, S.: Design of pattern recognition system for static security assessment and classification (2012).
10. Kozhirbayev, Z., Erol, B, A.: Sharipbay, A, A.: Speaker Recognition for Robotic Control via an IoT Device (2018).
11. Cui, L., Yang, S., Chen, F.: A survey on application of machine learning for Internet of Things. Int. J. Mach. Learn. & Cyber. 9, pp. 1399–1417 (2018).
12. Huskanovicm, A., Macan, A, A., Antolovic, Z., Tomas, B., Mijac, Marko.: Image pattern recognition using mobile devices (2013).
13. Fortino, G., Giordano, A., Guerrieri, A., Spezzano, G., Vinci, A.: A Data Analytics Schema for Activity Recognition in Smart Home Environments (2015).
14. Iver, R., Sharma, A.: IoT based Home Automation System with Pattern Recognition, vol. 8 (2019).
15. Raghavan, S., Tewolde, G, S.: Cloud based low-cost Home Monitoring and Automation System (2015).
16. Souza, A., Amazonas, J, R.: A Novel IoT Architecture with Pattern Recognition Mechanism and Big Data (2015).
17. R. Dave, "Utilizing Location Data and Enhanced Modeling Based on Daily Usage Pattern for Power Prediction and Consumption Reduction in Mobile Devices: A Low Power Android Application." North Carolina Agricultural and Technical State University, 2020.
18. Bhamare, D,. Suryawanshi, P,: Review on Reliable Pattern Recognition with Machine Learning Techniques (2018).
19. Iyer, D., Mohammad, F., Guo, Yuan.: Generalized Hand Gesture Recognition for Wearable Devices in IoT: Application and Implementation Challenges (2020).
20. Kim, E., Helal, S., Cook, D.: Human Activity Recognition and Pattern Discovery (2009).
21. Ketu, S., Mishra, P.: Performance Analysis of Machine Learning Algorithms for IoT-Based Human Activity Recognition (2020).
22. Peixoto, A., Vasconcelos, F., Guimaraes, M., Medeiros, A., Rego, P., Neto, A., Albuquerque, V., Filho, P.: A high-efficiency energy and storage approach for IoT applications of facial recognition, vol. 96 (2020)
23. Abomhara, M., Køien, G.: Cyber Security and the Internet of Things: Vulnerabilities, Threats, Intruders and Attacks, vol., 4, pp. 65-88 (2015).
24. Zainab, A., S. Refaat, S., Bouhali, O.: Ensemble-Based Spam Detection in Smart Home IoT Devices Time Series Data Using Machine Learning Techniques (2020).
25. Shi, C., Liu, J., Liu, H., Chen.: Smart User Authentication through Actuation of Daily Activities Leveraging WiFi-enabled IoT, (2017).
26. Ehatisham-ul-Haq, M., Azam, M.A.: Opportunistic sensing for inferring in-the-wild human contexts based on activity pattern recognition using smart computing, vol. 106, pp. 375-392 (2020).
27. Castro, D., Coral, W., Lopez, J.L.: Rodriguez, C., Colorado, J.: Wearable-Based Human Activity Recognition Using an IoT Approach (2017).
28. Al Safadi, E., Mohammad, F., Iyer, D., Smiley, B. J., Jain, N.K.: Generalized activity recognition using accelerometer in wearable devices for IoT applications, 2016 13th IEEE International Conference on Advanced Video and Signal Based Surveillance (2016).

29. Sarbishei, O.: A Platform and Methodology Enabling Real-Time Motion Pattern Recognition on Low-Power Smart Devices, 2019 IEEE 5th World Forum on Internet of Things, pp. 269-272, (2019).
30. Subasi, A., Radhwan, M., Kurdi, R., Khateeb, K.: IoT based mobile healthcare system for human activity recognition," 2018 15th Learning and Technology Conference (L&T), pp. 29-34 (2018).
31. Ehastisham-ul-Haq, M., Azam, M., Naeem, U., Amin, Y., Loo, J.: Continuous authentication of smartphone users based on activity pattern recognition using passive mobile sensing. Journal of Network and Computer Applications vol. 109, pp. 24-35 (2018).
32. Shoaib, M., Scholten, H., Havinga, P. J. M.: Towards physical activity recognition using smartphone sensors. In: 2013 IEEE 10th International Conference on Ubiquitous Intelligence and Computing and 2013 IEEE 10th International Conference on Autonomic and Trusted Computing, pp. 80-87. IEEE, Vietri sul Mere (2013).
33. Buriro, A., Crispo, B., Zhauniarovich, Y.: Please hold on: Unobtrusive user authentication using smartphone's built-in sensors. In: 2017 IEEE International Conference on Identity, Security and Behavior Analysis (ISBA), pp. 1-8. IEEE, New Delhi (2017).
34. Shafiq, M., Tian, Z., Sun, Y., Du, X., Guizani, M.: Selection of effective machine learning algorithm and Bot-IoT attacks traffic identification for internet of things in smart city. Future Generation Computer Systems vol. 107, pp. 433-442 (2020).
35. Lee, S., Wi, S., Seo, E., Jung, J.: ProFiOt: Abnormal behavior profiling (ABP) of IoT devices based on a machine learning approach. In: 2017 27th International Telecommunication Networks and Applications Conference (ITNAC), pp. 1-6. ResearchGate (2017).